\documentclass[prc,twocolumn,showpacs,preprintnumbers,amsmath,amssymb]{revtex4}

\usepackage{graphicx}

\usepackage{dcolumn}
\usepackage{bm}

\begin{document}
\bibliographystyle{/usr/share/texmf/tex/latex/revtex/prsty}
\draft
\title{
Potential relations for nonrelativistic and relativistic two-body
Schr\"odinger  equations
}

\author{
H. Kamada
\footnote{email:kamada@mns.kyutech.ac.jp}
}

\address{
Department  of Physics, Faculty of Engineering, Kyushu Institute of Technology,
 Kitakyushu 804-8550, Japan}

\author{
W. Gl\"ockle
\footnote{email:walterg@ruhr-uni-bochum.de}
}

\address{
Institut f\"ur  Theoretische Physik II,
Ruhr-Universit\"at Bochum, 44780 Bochum, Germany}

\date{\today}

\begin{abstract}
    Relations between nonrelativistic and relativistic two-body
equations, also allowing for different masses, are studied and explicit
expressions are given. One example is the Blankenbeclar Sugar equation.
The corresponding expressions for the boosted two-body potentials are
provided.
\end{abstract}

\maketitle

\narrowtext

The modern high precision NN potentials CD-Bonn \cite{CDBONN},
Nijm I,II \cite{NIJM} and
AV18 \cite{Argonne} 
  appear in a nonrelativistic Schr\"odinger equation. Turning to a
relativistic version thereof which replaces the nonrelativistic kinetic
energy by the relativistic one requires a modified NN potential to
describe the same NN phase shifts. In case of AV18 such
modifications have been worked out in \cite{ref4}. 
One can also use an analytical
momentum change transformation \cite{ref5}, 
which might be sufficient for a first
orientation , but which can not replace the necessary changes in the potential
introduced physics-wise ( see also \cite{ref6} for  more formal aspects).

 In case of CD-Bonn and the Nijmegen potentials relativistic kinematics
 has been applied relating the relative momentum $k$ and the energy $E$
  for two equal mass particles in the c.m.system.
 \begin{eqnarray}
 E= 2\sqrt{ m^2 + { \vec k } ^ 2 }.
\label{e1}
\end{eqnarray}
  For such a choice one can  switch  between  the Schr\"odinger equation in the
 relativistic  and  the nonrelativistic  form  by a simple trick
 \cite{coester,friar}. 
 Put $\omega = \sqrt{  m^2+k^2} $  then the relativistic NN
 Schr\"odinger equation reads
 \begin{eqnarray}
(2\omega + v ) \psi = 2 \omega _0 \psi .
\label{e2}
\end{eqnarray}
 Applying $( 2 \omega + v) $  again on both sides yields
 \begin{eqnarray}
 ( 4 \omega ^ 2 + 2 \omega v + 2 v \omega + v ^2 ) \psi = 4 \omega _0 ^2 \psi . 
 \label{e3}
 \end{eqnarray}
 After a simple algebra one arrives at the nonrelativistic Schr\"odinger
 equation
 \begin{eqnarray}
 ({ k ^2  \over m } + V_{nr} ) \psi = { k_ 0 ^2 \over m } \psi
 \label{e4}
 \end{eqnarray}
 with
 \begin{eqnarray}
V_{nr} \equiv { 1 \over { 2 m }} ( \omega v + v \omega + { 1 \over 2 } 
v ^2 ) .
\label{e5}
 \end{eqnarray}
 This means that solving Eq.(\ref{e4}) together with the relativistic kinematic
 relation (\ref{e1})  is in fact a relativistic treatment. One can
 easily invert
 the steps from Eq.(\ref{e2}) to (\ref{e5}) and finds
 \begin{eqnarray}
v = \sqrt{4m V_{nr} + 4 \omega ^2  } - 2 \omega
\label{e6}
  \end{eqnarray}
 which relates  the  potential $v$  in the relativistic NN
 Schr\"odinger equation  to the potential  $V_{nr}$ in the nonrelativistic
 Schr\"odinger equation. Important thereby is of course the relation
 (\ref{e1}).
 This connection is also valid for the Nijmegen potentials, since
 also there the relativistic relation (\ref{e1}) has been used.

 In previous work ( see for instance \cite{keiser91, gloeckle86})
it is stated that the bound state eigenvalue to Eq. (\ref{e4}) , which
is $M_b^2/4m$ - m differs
from $E_b  \equiv  M_b-2m$ by the small amount $E_b^2/
4m$. Here $M_b$ is the mass of the
bound two-body system. Though this is correct one can use a different
definition of the
binding energy, which replaces  $E_b$ by $\epsilon_b$ defined as
\begin{eqnarray}
     M_b \equiv \sqrt{ 4m^2 + 4 \epsilon_b m} 
\label{eq7}
\end{eqnarray} 
Then in Eq.(\ref{e4}) $\epsilon_b$ occurs directly without correction
term and moreover it can
naturally be written as
 $  \epsilon_b = - \kappa^2/m $
in agreement with the form of the energy eigenvalue in
Eq.(\ref{e4}). Expanding Eq.(\ref{eq7}) one obtains
$ 
M_b= 2m+\epsilon_b - \epsilon_b^2/4m +...
$ 
the lowest order of which agrees with the usual definition of the
binding energy. 

  For the practical application of (\ref{e6}) one can proceed in close
  analogy to 
  the representation shown in \cite{ref9} for the boosted NN potentials. We
  obtain( for details see Appendix )
\begin{eqnarray}
&&\langle \vec k \vert v  \vert \vec k' \rangle
 =  \Psi_b(\vec k)   M_b  \Psi_b (\vec k')
 \cr &+& {   m \over { k ^2 - {k '}^2 } } \{  2 \omega \Re 
[ T (\vec k' , \vec k ;
 { {k}^2 \over m})]
 -  2 \omega' \Re [ T (\vec k , \vec k' ; { {k'}^2 \over m} )] \}
 \cr &+&  {m^2 \over {  k ^2 - {k '}^2 } }\times
 \cr 
 & \{ &  {\cal P}  \int d^3 k''  { 2  \omega''  \over {{ { k ''} ^2 - {k }^2 }} }
  T (\vec k , \vec k'' ; { {k''}^2 \over m}) 
  T^{*} (\vec k' , \vec k'' ;  { {k''}^2 \over m} )
   \cr &-& {\cal P}
 \int d^3 k''{  2  \omega''  \over { { k ''} ^2 - {k '}^2 } }
   T (\vec k , \vec k'' ;{ {k''}^2 \over m} ) 
   T^{*} (\vec k' , \vec k'' ; { {k''}^2 \over m} ) 
    \}.
    \label{e7}
    \end{eqnarray}
  where $\Psi_b (k)$ is the  nonrelativistic deuteron wave function of
    Eq(4), $M_b$ the mass of the
  deuteron, $T$ the standard NN T-matrix related to $V_{nr}$ via the
    nonrelativistic Lippmann Schwinger equation and $\omega '' =
    \sqrt{m^2 +k''^2}$.

   Like in \cite{ref9}
   one can also  obtain the boosted potential $v_p$ in a frame, where the
   total  NN momentum is different from zero:
\begin{eqnarray}
v_p = \sqrt{ ( 2 \omega + v ) ^2 + { \vec p }~ ^2 } - \sqrt{ 4 \omega ^2 + 
{ \vec p }~ ^2 }.
\label{e8}
\end{eqnarray}
   In this case $ v _p$ is connected to $V_{nr}$ as
\begin{eqnarray}
v_p = \sqrt{ 4 m V_{nr}+4 \omega ^2  + { \vec p }~ ^2 } - \sqrt{ 4 \omega ^2 +
{ \vec p }~ ^2 }.
\label{e9}
\end{eqnarray}
The explicit representation is
\begin{eqnarray}
&&\langle \vec k \vert v_p  \vert \vec k' \rangle
 =  \Psi_b(\vec k)  \sqrt{   {M_b} ^2 + {\vec p }~^2 }  \Psi_b (\vec k')
  \cr &+& {  m \over { k ^2 - {k '}^2 } } 
  \{ \sqrt{4  \omega ^2  + {\vec p }~^2 }
  \Re [ T (\vec k' , \vec k ;
   { {k}^2 \over m})]
\cr
&-&  \sqrt{4 { \omega '} ^2  + {\vec p }~^2 }
  \Re [ T (\vec k , \vec k' ; { {k'}^2 \over m} )] \}
     \cr 
     &+&  { m^2 \over {  k ^2 - {k '}^2 } }\times
      \cr
& \{ &  {\cal P}  \int d^3 k'' {\sqrt{4 {\omega ''} ^2 + {\vec p }~^2 } 
  \over {{ k ''} ^2 - {k }^2 } }
         T (\vec k , \vec k'' ; { {k''}^2 \over m})
           T^{*} (\vec k' , \vec k'' ;  { {k''}^2 \over m} )
    \cr &-&  {\cal P} \int d^3 k''  {  \sqrt{4 {\omega ''} ^2 + {\vec p }~^2 }
    \over { { k ''} ^2 - {k '}^2 } }
 T (\vec k , \vec k'' ;{ {k''}^2 \over m} )
    T^{*} (\vec k' , \vec k'' ; { {k''}^2 \over m} )
                        \}.
                            \label{e10}
\cr &&
                                \end{eqnarray}
This is to be used together with the kinetic energy $h_0 = \sqrt{ 4 \omega
^2  + { \vec p ~} ^2 }$. 

In \cite{ref9} we did not use the correct relation 
(\ref{e6}) for CD-Bonn and the
    Nijmegen
    potentials to arrive at $v$ but the analytical momentum change
    transformation [5]. Using (\ref{e6})  might change the outcome for the
    relativistic energy shifts in the triton binding energy shown in 
    \cite{ref9}.
    This is left to a forthcoming calculation.

Now we ask a more general question. Assume as a starting point a
relativistic three- dimensional equation for a two-body system
with  general masses in the total momentum zero frame, which defines the
wave functions for bound and scattering states. It is assumed that they
fulfill the completeness relation
\begin{eqnarray}
\sum _ b \vert \Psi_b > < \Psi_b \vert + \int d \vec k \vert \Psi _{\vec k }
>< \Psi_{\vec k} \vert = 1 
\label{e11}
\end{eqnarray}
and obey the homogeneous and inhomogeneous equations
\begin{eqnarray}
\vert \Psi_b > = G_0 V \vert \Psi_b >,
\label{e12}
\end{eqnarray}
\begin{eqnarray} 
\vert \Psi _{\vec k }> = \vert \vec k > + G_0 T \vert \vec k>.
\label{e13}
\end{eqnarray}
Here  $V$, $G_0$ and $T$ are the corresponding potential, free propagator
( with the  correct  kinematical cut) and T-operator, respectively. 
( Note , this $T$ is different from the previous one). 
Then
the energies related to $\vert \Psi_b >$
are $M_b$ and for  the scattering states $ \vert \Psi _{\vec k }>$
they are $\omega _ 1 +\omega _2 $ with 
$\omega_i = \sqrt{m_i ^2 + {\vec k~}^2}$. We want to rewrite these
equations into the form of a standard relativistic two-body equation in the
system  with total momentum zero:
\begin{eqnarray}
h \equiv \hat \omega _1 + \hat \omega _2 + v 
\label{e14}
\end{eqnarray}
 where $\hat \omega _i $ are the corresponding operators related to $\omega_i$. This means that $h$
 should have the same bound and scattering states. Therefore  the
 operator $h$ can be represented in terms of the bound and
 scattering states of the underlying relativistic two-body equation as
\begin{eqnarray}
h&=&\sum_b \vert  \Psi_b > M_b  < \Psi_b \vert
 \cr &+& 
 \int d \vec k \vert \Psi _{\vec k }
 > ( \omega_1  ( \vec k ) +  \omega_2  ( \vec k ) )  < \Psi_{\vec k}
 \vert .
\label{e15}
\end{eqnarray}
 Using that form one obviously gets
\begin{eqnarray}
v   & \equiv & h - \hat \omega _1 - \hat \omega _2 
=
\sum_b \vert \Psi_ b >  M_b < \Psi_b \vert \cr 
 &+& \int d \vec k \vert 
 \Psi _{\vec k} >  ( \omega_1  ( \vec k ) +  \omega_2  ( \vec k ) )
 < \Psi _{\vec k} \vert \cr
 &-& \int d \vec k \vert \vec k >
   ( \omega_1  ( \vec k ) +  \omega_2  ( \vec k ) ) 
 < \vec k \vert .
\label{e16}
\end{eqnarray}
 This leads to the momentum representation
\begin{eqnarray}
&&< \vec k \vert v \vert {\vec k } ' > = \sum _b 
\Psi_b  (\vec k ) M_b \Psi_ b ( { \vec k } ') 
\cr &+&
\int d \vec k '' < \vec k \vert \Psi _ { \vec k ''} > 
 ( \omega_1  ( \vec k '' ) +  \omega_2  ( \vec k '') ) 
 < \Psi _ { \vec k ''} \vert \vec k '> 
\cr &-&
 \delta ( \vec k - \vec k ' )  ( \omega_1  ( \vec k ) +  \omega_2  ( \vec k ) ).
\label{e17}
\end{eqnarray}
 We insert now the underlying form of the assumed relativistic equation (\ref{e13})
 and get for the part with the scattering states
\begin{eqnarray}
&&< \vec k \vert v \vert \vec k ' > _{scatt} 
\cr  &=& 
\int d \vec {k''}
\{ \delta (\vec k - \vec k'' ) + G_0 ( k, k'' ) < \vec k \vert T \vert \vec k ''> \}
\cr && 
(  \omega_1  ( \vec k '' ) +  \omega_2  ( \vec k '') )
\{ < \vec k ' \vert T \vert \vec k '' > ^* G _0 ^* (k', k'') + \delta ( \vec k'' -
\vec k' ) \} \cr
&&-\delta(\vec k -\vec k') ( \omega_1  ( \vec k  ) +  \omega_2  ( \vec k ) )
.
\label{e18}
\end{eqnarray}
 As an example we regard the Blankenbeclar-Sugar(BBS) equation resulting from
 the Bethe-Salpeter (BS) equation by replacing the propagator in the BS
 equation with the propagator \cite{ref10}
\begin{eqnarray}
&&g_{BBS} \equiv { 1 \over { 2 \pi i } } \int _ { (m_1 + m_2)^2 } ^\infty d s ' 
i ( 2 \pi ) ^2 { \delta ^+ ( k_1 ^2 - m_1 ^2 ) \delta^+ ( k_2 ^2 - m_2 ^2 ) 
\over s ' - s - i \epsilon }.  
\label{e19}
\end{eqnarray}

 Putting the four- momenta $k_{1,2} = { 1 \over 2} P \pm k $
  and   $P_0 $($P'_0$)$ =  \sqrt{s}$ ( $ \sqrt{s'}$ )
  one obtains \cite{ref10}
\begin{eqnarray}
g_{BBS}= { \pi~~  \delta ( k_0 - { 1 \over 2 }
(\omega_1 - \omega _2) )  \over  \omega _ 1 \omega _2 }
{ \omega _ 1 + \omega _2 \over ( \omega _1 + \omega _2 )
^2 - s - i \epsilon }. \cr
\label{e20}
\end{eqnarray}

   This yields the following equation of the type (\ref{e13}):
\begin{eqnarray}
   |\Psi_{\vec k' }> = | \vec k'> + { 1 \over  ( \hat \omega_1 + \hat \omega_2 )^2 - ( \omega_1' +
\omega_2')^2 - i \epsilon }  T | \vec k'>
\label{e21}
\end{eqnarray}
In arriving at (\ref{e21}) we have redefined the matrix elements of $T$ and $V$
occurring in the Bethe Salpeter equation after inserting (\ref{e20}) by
multiplying from both sides $\sqrt{ \pi ( \omega_1+\omega_2)/\omega_1
\omega_2}$ and $\sqrt{ \pi ( \omega_1' + \omega_2') / \omega_1' \omega_2' }$,
respectively.
We denote $ \omega_i' = \sqrt{m_i^2+ {k'}^2 }$.

 The bound states are defined by an equation corresponding to (\ref{e12}) with
the same propagator as in (\ref{e21}) and the redefined $V$. We assume the
completeness relation(\ref{e11}) to be valid and obtain the two-body
interaction
$v$ in Eq.(\ref{e14}) as
\begin{eqnarray}
&&< \vec k \vert v \vert \vec k ' > = \sum_b
 \Psi_b(\vec k)   M_b  \Psi_b (\vec k')
 \cr &+& 
{  1  \over {  ( \omega _1 + \omega _2 )
^2 - ( \omega _1'  + \omega _2'   )
^2  } } 
 \times \cr  && 
\{ ( \omega _1 + \omega _2 )   \Re [ T (\vec k' , \vec k ;
  \omega _1 + \omega _2   )] 
\cr &&
 -  (\omega_1' + \omega _2')  \Re [ T (\vec k , \vec k' ; 
 \omega _1' + \omega _2 ' )] \}
 \cr &+&  
{  1  \over {  ( \omega _1 + \omega _2 )
^2 - ( \omega _1'  + \omega _2'   )
^2  } } 
\times
 \cr 
 & \{ &  {\cal P}  \int d^3 k'' 
 { \omega _1 '' + \omega _2 ''  \over
 {  ( \omega _1'' + \omega _2'' )
^2 - ( \omega _1  + \omega _2  )
^2  }  } \times
\cr &&
  ~T (\vec k , \vec k'' ; {  \omega _1'' + \omega _2''  }) 
  ~T^{*} (\vec k' , \vec k'' ;  {  \omega _1'' + \omega _2''} )
\cr &-& 
 {\cal P}  \int d^3 k'' 
 { \omega _1 '' + \omega _2 ''  \over
 {  ( \omega _1'' + \omega _2'' )
^2 - ( \omega _1'  + \omega _2'  )
^2  }  } \times
\cr &&
  ~T (\vec k , \vec k'' ; {  \omega _1'' + \omega _2''  }) 
  ~T^{*} (\vec k' , \vec k'' ;  {  \omega _1'' + \omega _2''} )
    \}.
\label{e22}
\end{eqnarray}

  To arrive at (\ref{e22}) the steps are analogous to the steps laid out in
the Appendix.

  The boosted potential which is to be taken together with the kinetic
energy $\sqrt{ ( \omega_1 + \omega_2)^2 + p^2}$ results simply be replacing
$M_b$ by $\sqrt{ M_b^2 + p^2}$ and the $( \omega_1 + \omega_2)$'s in the
numerators by $\sqrt{ ( \omega_1 + \omega_2)^2 +p^2}$.

 Finally we address the question whether the relativistic two-body
 Schr\"odinger equation corresponding to $h$ from Eq(\ref{e14}) can be cast into the
 form of a nonrelativistic two-body Schr\"odinger equation
\begin{eqnarray}
h_{nr} \Psi \equiv ( {{ \hat {\vec k} }~ ^2  \over 2 \mu } + v _{nr} ) \Psi = {{ \vec {k} }~ ^2  \over 2 \mu }  \Psi 
\label{e23}
\end{eqnarray}
 where $\mu$  is the standard reduced mass
( $ { 1 \over \mu}  = { 1 \over m_1 } + { 1 \over m_2 }.
$)
 We require again that  $h_{nr} $  has the same bound and scattering states as
 $h$. Therefore
\begin{eqnarray}
&&v_{nr} \equiv h_{nr } - {\hat {\vec k } ^2 \over 2 \mu } 
= \sum _b \vert \Psi _b > \epsilon _b < \Psi _b \vert 
\cr
&&+ \int d \vec k \vert \Psi _{ \vec k } > { { \vec k }^2 \over 2 \mu }
 < \Psi _{ \vec k } \vert  -
{ \hat   { \vec k }^2 \over 2 \mu }.
\label{e26}
 \end{eqnarray}
In momentum representation this yields
\begin{eqnarray}
&& < \vec k \vert v_{nr} \vert \vec k ' > =
\sum _b \Psi _b (\vec k) \epsilon_b  \Psi _b (\vec k ')  
\cr
&& + \int d \vec k '' < \vec k \vert \Psi _{ \vec k'' } > 
 { { \vec {k ''} }^2 \over 2 \mu }
< \Psi _{ \vec k ''} \vert \vec k '> \cr  
&& - \delta ( \vec k  - \vec k ') {   { \vec k }^2 \over 2 \mu }
\cr &&
 = \sum_b
 \Psi_b(\vec k)   \epsilon _b  \Psi_b (\vec k')
 \cr &+& 
{  1  \over {  ( \omega _1 + \omega _2 )
^2 - ( \omega _1'  + \omega _2'   )
^2  } }    \times \cr  && 
\{   { { \vec {k } }^2 \over 2 \mu }   \Re [ T (\vec k' , \vec k ;
  \omega _1 + \omega _2   )]  - { { \vec {k '} }^2 \over 2 \mu }
  \Re [ T (\vec k , \vec k' ; 
 \omega _1' + \omega _2 ' )] \}
 \cr &+&  
{  1  \over {  ( \omega _1 + \omega _2 )
^2 - ( \omega _1'  + \omega _2 '  )
^2  } } 
\times
 \cr 
 & \{ &  {\cal P}  \int d^3 k'' 
 {  { { \vec {k ''} }^2 \over 2 \mu } \over
 {  ( \omega _1'' + \omega _2'' )
^2 - ( \omega _1  + \omega _2  )
^2  }  }
\cr &&
  T (\vec k , \vec k'' ; {  \omega _1'' + \omega _2''  }) 
  ~T^{*} (\vec k' , \vec k'' ;  {  \omega _1'' + \omega _2''} )
\cr &-& 
 {\cal P}  \int d^3 k'' 
 {  { { \vec {k ''} }^2 \over 2 \mu }  \over
 {  ( \omega _1'' + \omega _2'' )
^2 - ( \omega _1'  + \omega _2'  )
^2  }  }
\cr &&
  T (\vec k , \vec k'' ; {  \omega _1'' + \omega _2''  }) 
  ~T^{*} (\vec k' , \vec k'' ;  {  \omega _1'' + \omega _2''} )
    \}.
\label{e27}
 \end{eqnarray}
 In summary we regarded relations between relativistic and
nonrelativistic two-body equations. For equal mass particles the
relativistic two-body Schr\"odinger  equation can be converted
identically into a nonrelativistic Schr\"odinger equation and vice versa
if the relativistic connection between energy and momentum is kept. An
explicit expression for the relativistic potential in terms of the
nonrelativistic T-matrix and bound state is provided. Also the
corresponding relativistic potential in a moving frame has been worked
out. Further we regarded a more general question. Often relativistic
three-dimensional equations are proposed, which do not have the standard
form of a relativistic two-body Schr\"odinger equation. Can they be
rewritten into such a form? We addressed that question and examplified a
solution for the case of the Blankenbeclar Sugar equation for two
different mass  particles. Finally we provided an explicit expression for
the nonrelativistic potential occurring together with the standard
nonrelativistic kinetic energy of two different mass particles if bound
and  scattering state information is available from the underlying
relativistic two-body Schr\"odinger equation. In this case
because of two square root expressions in the kinetic energy the simple
trick possible for two equal mass particles is not applicable and we are
left with a more complicated form.

One of authors (H.K.) would like to thank  the Deutsche
Forschungsgemeinschaft for financial support.

\bigskip


{\bf 
Derivation of Eq.(\ref{e7}):}

\noindent
We insert the complete basis of eigenstates related to Eq. (\ref{e6}) 
(Assuming  one bound state). This leads to
\begin{eqnarray}
&&< \vec k \vert v \vert \vec k~ '>= < \vec k \vert \Psi_b> M_b< \Psi_b \vert \vec k ~' > 
- 2 \omega \delta(\vec k - \vec k') \cr  
&+& \int d \vec k '' < \vec k \vert \Psi_{\vec k''} > \sqrt{4m} \sqrt{ 
{k''^2 \over m} +m }< \Psi_{\vec k''}  \vert \vec k ~'>
\label{a1}
\end{eqnarray}
Next we use the well known decomposition
\begin{eqnarray}
 < \vec k \vert \Psi_{\vec k'} > = \delta (\vec k - \vec k')+ { T(\vec
 k,\vec k'; {k'^2 \over m }) \over {k'^2 \over m }+ i \epsilon -  {k^2
 \over m } } 
\label{a2}
\end{eqnarray}
and arrive at 
\begin{eqnarray}
&&< \vec k \vert v \vert \vec k~ '>= 
 \Psi_b (\vec k)  M_b \Psi_b (\vec k ~')  \cr  
&+& { T^*(\vec
 k',\vec k; {k^2 \over m }) \over {k^2 \over m }- i \epsilon -  {k'^2
 \over m } } 2 \omega( k)
+{ T^(\vec
 k,\vec k'; {k'^2 \over m }) \over {k'^2 \over m }+ i \epsilon -  {k^2
 \over m } } 2 \omega( k') \cr 
&+&\int d \vec k ''  { T^(\vec
 k,\vec k''; {k''^2 \over m }) \over {k''^2 \over m }+ i \epsilon -  {k^2
 \over m } } 2 \omega(\vec k'')
  { T^*(\vec
 k',\vec k''; {k''^2 \over m }) \over {k''^2 \over m }- i \epsilon -  {k'^2
 \over m } }
\label{a3}
\end{eqnarray}
The integral requires some care and we keep the limiting processes for
the two scattering states separately by putting 
\begin{eqnarray}
&&
{1 \over k''^2 - k ^2 +i \epsilon }
{1 \over k''^2 - {k' } ^2 -i \epsilon } \cr &\to& 
{1 \over k''^2 - k ^2 +i \epsilon_1 }
{1 \over k''^2 - {k' } ^2 -i \epsilon_2 }
\cr
&=& \left( {1 \over k''^2 - k ^2 +i \epsilon_1 } - 
{1 \over k''^2 - {k' } ^2 -i \epsilon_2 } \right) \times 
\cr &&
{1 \over k^2 - {k' } ^2 -i( \epsilon_1 + \epsilon_2)  } .
\label{a4}
\end{eqnarray}
This allows us to perform one limit firstly with the result
\begin{eqnarray}
&&
{1 \over k''^2 - k ^2 +i \epsilon }
{1 \over k''^2 - {k' } ^2 -i \epsilon } \cr &\to& 
\left( { {\cal P  } \over   k''^2 - k ^2 } - i \pi \delta ( k''^2 - k
^2 ) \right) {1 \over k^2 - {k' } ^2 -i \epsilon_2  } \cr 
&-&
\left( { {\cal P  } \over   k''^2 - {k' } ^2 } + i \pi \delta ( k''^2 -{k' } 
^2 ) \right) {1 \over k^2 - {k' } ^2 -i \epsilon_1  } 
\label{a5}
\end{eqnarray}
Thus we get for some well behaved function $f (\vec k '') $
\begin{eqnarray}
&&
\int d {\vec k '' } { f (\vec k '') \over ( k''^2 - {k } ^2 +i
\epsilon_1 ) 
(k''^2 - {k' } ^2 -i \epsilon_2)  } \cr
&&
= \lim _{\epsilon \to +0  } {1 \over k^2 - {k' } ^2 -i \epsilon } \times
\cr &&
\left( {\cal P } \int d { \vec k '' } {f(\vec k '' ) \over k''^2 - k^2  }
-  {\cal P } \int d { \vec k '' } {f(\vec k '' ) \over k''^2 -{ k'}^2
}
\right) \cr 
&&- i\pi \lim_{\epsilon \to +0  } {1 \over k^2 - {k' } ^2 -i \epsilon } \times
\cr &&
\left( \int d { \vec k '' } f(\vec k '' ) \delta ( k''^2 - k^2 )
+ \int d { \vec k '' } f(\vec k '' ) \delta (k''^2 -{ k'}^2 )
\right)
\cr && 
\label{a6}
\end{eqnarray}
The principal value prescription is denoted via ``${\cal P}\int$''. 
In our case
\begin{eqnarray}
f(\vec k '') =
2m^2 \omega (k'') T(\vec k, \vec k'' ; { k''^2 \over m }) T^*(\vec k', 
\vec k'' ; { k''^2 \over m }) 
\label{a7}
\end{eqnarray}
and therefore
\begin{eqnarray}
&&\int d\vec k '' f(\vec k'') \delta ( k''^2 - k^2 ) \cr
&=&2m^2 \int d \vec k '' \omega(k'')  T(\vec k, \vec k'' ; { k''^2
\over m }) T^*(\vec k', 
\vec k'' ; { k''^2 \over m })
\cr && \delta ( k''^2 - k^2 )
\label{a8}
\end{eqnarray}
This is part of the unitary relation 
\begin{eqnarray}
&&
 T(\vec k, \vec k'' ; { q^2 \over m }) - T^*(\vec k', 
\vec k'' ; { q^2 \over m })
\cr &=& 
-2 i \pi m \int d \vec k''  T(\vec k, \vec k'' ; { q^2 \over m }) T^*(\vec k', 
\vec k'' ; { q^2 \over m }) \delta( q^2 - k''^2) 
\cr &&
\label{a9}
\end{eqnarray}
Consequently 
\begin{eqnarray}
&& 2m^2 \int d\vec k'' \omega(k '')  T(\vec k, \vec k'' ; { k''^2
 \over m })
 T^*(\vec k', 
\vec k'' ; { k''^2 \over m })  \delta (k''^2 - k^2) 
\cr &=& 
- {2 \omega(k) m \over \pi  }
\Im [ T(\vec k , \vec k' ; { k ^2 \over m } ) ] 
\label{a10}
\end{eqnarray}
and 
\begin{eqnarray}
&& -i\pi \lim_{\epsilon \to +0  } { 1 \over k^2 - {k' }^2  -i \epsilon
} \times  
\cr && 
\left( \int d \vec k '' f(k'') \delta (k''^2 - k^2) +
\int d \vec k'' f(k'') \delta ( k''^2 - {k' }^2 ) \right)\cr 
&&=\lim _{\epsilon \to +0  } { 2 m \over k^2 -{k'}^2-  i \epsilon } \times 
\cr && 
\left( 
\omega (k) \Im [T ( \vec k, \vec k' ; { k^2 \over m  } )] +
\omega (k') \Im [T (\vec k, \vec k' ; { {k'}^2 \over m })]  \right)
\label{a11}
\end{eqnarray}
Combined with Eq. (\ref{a3}) certain terms cancel and one arrives at
Eq.(\ref{e7}).

\end{document}